\documentclass[12]{article}
\usepackage{amsmath, amssymb}
\begin{document}
\author{Johan Noldus, Vakgroep Wiskundige Analyse, \\* Universiteit Gent, Galglaan 2, 9000 Gent, Belgium}

\title{A Lorentzian Gromov-Hausdorff notion of distance}
\maketitle
\begin{abstract}
This paper is the first of three in which I study the moduli space of isometry classes of (compact) globally hyperbolic spacetimes (with boundary).  I introduce a notion of Gromov-Hausdorff distance which makes this moduli space into a metric space.  Further properties of this metric space are studied in the next papers.  The importance of the work can be situated in fields such as cosmology, quantum gravity and - for the mathematicians - global Lorentzian geometry.      
\end{abstract}
\maketitle
\section{Introduction}  The aim of this paper is to make first steps in the construction of a convergence theory for partially ordered spaces equipped with a Lorentz distance, which we shall refer to as Lorentz spaces.  Typical examples of such spaces are Lorentz manifolds, which constitute the geometrical playground of general relativity.  The field of application should in the end be quantum gravity and in particular the path integral formulation thereof.  In this application, the purpose is twofold:  on one hand a convergence theory will serve as a tool for taking a continuum limit, on the other hand it will provide a mechanism to control which geometrical objects to sum over and which not.  It is my hope that in a later stage, we shall be able to link this rather abstract control theory with statistical Lorentzian geometry, in order to be able to make this passage to the former application.  \\*
At the moment, the main background-independent attempts to quantize general relativity are canonical quantum gravity and the resulting spin foam models, which are structural extensions of the causal sets introduced by Rafael Sorkin.  The main difficulty in all these approaches consists in making precise what it means for such a Lorentz space (in fact, causal sets are not Lorentz spaces in the above sense since the Lorentzian distance is not accounted for) to be close to a Lorentz manifold.  Other difficult questions with respect to these objects concern a good definition of dimensionality and relativistic scale.  In these three papers we shall present an \emph{abstract} solution to all these problems in the way Gromov did for locally compact metric spaces.  The Lorentzian analogue is not just a copy of the $23$ year old Gromov theory \cite{Gromov} \cite{Petersen}: some intermediate results have to be stated differently and the proofs are considerably more difficult. \\*
In section two, we generalize the Lipschitz notion of distance between two compact metric spaces to a notion of distance between two globally hyperbolic compact Lorentz manifolds with spacelike boundary.  It turns out that we can only measure a distance between conformally equivalent structures, and in this sense this chapter is only a warm-up.  
The most important result is a Lorentzian analogue for the Ascoli-Arzela theorem which guarantees convergence to isometry.  The most important lesson, however, is that we have found a class of mappings that is rich enough to compare conformally equivalent structures and which is poor enough to keep a good control over.  
In section three we introduce a Lorentzian notion of Gromov - Hausdorff distance $d_{\textrm{GH}}$.  I will give some examples to show that the Lorentzian theory is rather different from the ``Riemannian'' one.  \\* Previous attempts in the literature to construct a metric on the modulo space of isometry classes of \emph{Lorentzian} spacetimes can be found in \cite{Noldus}, \cite{Bombelli} , \cite{Bombelli2} , \cite{Bombelli3}.  But all these attempts failed since one could only prove that one had obtained a \emph{pseudo} distance.  Moreover, I do not agree with the philosophy behind them, since in all these papers (including mine) the canonical volume measure has been used.  In particular, this means that I take the point of view that the construction of a \emph{statistical} Lorentzian convergence theory should follow a \emph{geometrical} Lorentzian convergence theory and not the other way around.  \\* 
The readers not familiar with the following notions and results concerning causality are, if not mentioned otherwise referred to the bible of general relativity, \cite{Hawking1}.

\section{A Lipschitz distance}
Our aim is to define a ``Lorentzian'' analogue of the classical ``Riemannian'' Lipschitz distance between (locally) compact (pointed) metric spaces.  Let $(X,d_{X})$ and $(Y,d_{Y})$ be two compact metric spaces and $f : X \rightarrow Y$ be a bi-Lipschitz mapping, i.e., there exist numbers $ 0 < \alpha < \beta $ such that $$ \alpha d_{X} (x,y) \leq d_{Y} (f(x),f(y)) \leq \beta d_{X} (x,y) \quad \forall x,y \in X $$
The minimal such $\beta$ is the \emph{dilatation} $\textrm{dil}(f)$ of $f$ and the maximal such $\alpha$ the \emph{co-dilatation} of $f$ ( or the inverse of the dilatation of $f^{-1}$ if $f^{-1}$ exists).  The Lipschitz distance $d_{L}(X,Y)$ between $X$ and $Y$ is the infimum over all bi-Lipschitz homeomorphisms of the expression: $$ \left| \ln (\textrm{dil}(f)) \right|  + \left| \ln( \textrm{dil}(f^{-1}) ) \right| $$
The key result is that $d_{L}(X,Y) = 0$ iff $(X,d_{X})$ is isometric to $(Y,d_{Y})$, which is a direct consequence of the Ascoli-Arzela theorem \cite{Yosida}.    
\newtheorem{theo}{Theorem}
\begin{theo} (\textbf{Ascoli-Arzela})
Let $X$ and $Y$ be second countable, locally compact spaces, moreover $(Y,d_{Y})$ is assumed to be metrically complete.  Assume that the sequence $\left\{ f_{n} \right\}$ of functions $f_{n} : X \rightarrow Y $ is equicontinuous such that the sets $ \bigcup_{n} \left\{f_{n}(x) \right\}$ are bounded with respect to $d_{Y}$ for every $x \in X$.  Then there exists a continuous function $f : X \rightarrow Y$ and a subsequence of $\left\{ f_{n} \right\}$ which converges uniformly on compact sets in $X$ to $f$.
\end{theo}
Let us now make some analogy and discrepancy with the Lorentzian case.  For now we restrict to spacetimes, i.e., pairs $(\mathcal{M},g)$ where  $\mathcal{M}$ is a $C^{\infty}$, paracompact, Hausdorff manifold and $g$ is a Lorentzian metric tensor on it, such that $(\mathcal{M},g)$ is time orientable.  Abstract Lorentzian spaces will be defined later on in analogy with Seifert and Busemann.  We will make now a convention in terminology which is not standard in the literature, but is somehow necessary to keep the discussion clear.
\newtheorem{deffie}{Definition}
\begin{deffie}
Let $X$ be a set, a Lorentzian distance is a function $d: X \times X \rightarrow \mathbb{R}^{+} \cup \left\{ \infty \right\} $ which satisfies
\begin{itemize}
\item $d(x,x) = 0$ 
\item $d(x,y) > 0$ implies $d(y,x) = 0$ (antisymmetry) 
\item if $d(x,y)d(y,z) > 0$ then $d(x,z) \geq d(x,y) + d(y,z)$ (reverse triangle inequality)
\end{itemize}
\end{deffie} 
It is well known that every chronological spacetime determines a canonical Lorentzian distance by defining $d_{g} (x,y)$ as the supremum over all lengths of future oriented causal curves from $x$ to $y$ if such curves exist and zero otherwise.  One has that $d_{g}$ is continuous and finite if $(\mathcal{M},g)$ is globally hyperbolic and, vice versa, that if $d_{g}$ is continuous then $(\mathcal{M},g)$ is causally continuous.  More equivalences between properties of $d_{g}$ and causality restrictions can be found in \cite{Beem}.  We shall only be interested in globally hyperbolic spacetimes since the continuity of $d_{g}$ is a desirable property if one wants to work out a comparison theory according to Lipschitz.  Note immediately that a compact globally hyperbolic spacetime does not exist unless we consider manifolds with a boundary.  We assume the boundary is spacelike and that $\mathcal{M}$ and $\mathcal{N}$ are locally extendible across their boundary (all the results in this paper are also valid when an extra timelike or null boundary is allowed, it is up to the reader to fill in the details in the proofs).  Remark first that every point of the boundary is contained in a neighborhood $\mathcal{U}$ which is diffeomorphic to a hypercube in $\mathbb{R}^{n}$ which is closed on one face and otherwise open.  By local extendibility I mean that there exists an isometric embedding of $(\mathcal{U}, g_{| \mathcal{U}})$ in a open spacetime $(\mathcal{V} , g_{ | \mathcal{V}})$ such that the image of $\mathcal{U}$ has compact closure in $\mathcal{V}$.  We stress that this does not correspond to the usual notion of \emph{causal} local extendibility to which we come back later on.\\*
First we have to contemplate which mappings between two spacetimes need to be considered for comparison.  To this purpose, let $(\mathcal{M},g)$ and $(\mathcal{N},h)$ denote globally hyperbolic spacetimes.  A mapping $f : \mathcal{M} \rightarrow \mathcal{N}$ is said to be timelike Lipschitz if and only if it has bounded timelike dilatation $\textrm{tdil} (f)$, i.e., there exists a (smallest) number $\beta$ such that 
$$ d_{h} ( f(x),f(y)) \leq \beta d_{g} (x , y), \quad \forall x,y \in \mathcal{M}. $$
The above construction for the ``Riemannian'' case suggests that we consider timelike bi-Lipschitz homeomorphisms.  However, a slight generalization of a classical result for homothecies teaches us that a surjective timelike bi-Lipschitz map is automatically a homeomorphism (see Appendix A).  Indeed, a result of Hawking, Mc Carthy and King \cite{Hawking2}, proves that such mapping is a $C^{\infty}$ conformal diffeomorphism.  One might be concerned that such maps are too restrictive in the sense that they only allow conformally equivalent spacetimes to be compared but as mentioned before, this section is meant as a warm-up to get used to the techniques needed for the next section.  The next logical step is formulating and proving a Lorentzian version of a suitably modified Ascoli-Arzela theorem.  
\begin{theo} (\textbf{Lorentzian Ascoli-Arzela}) Let $f_{n} : \mathcal{M} \rightarrow \mathcal{N}$ be onto bi-Lipschitz mappings such that $\bigcup_{n} \left\{ f_{n} (x) \right\}$ and $\bigcup_{n} \left\{ f^{-1}_{n} (y) \right\}$ are precompact\footnote{A subset $A$ of a topological space $X$ is precompact iff the closure of $A$ is compact.} in $\mathcal{N}$ respectively $\mathcal{M}$ for all $x \in \mathcal{M}$ and $y \in \mathcal{N}$.  Moreover, let $(c_{n})_{n \in \mathbb{N}}$ be a descending sequence ($c_{n} < 1$) converging to zero such that $\textrm{tdil} (f_{n}) \leq 1+c_{n}$ and $\textrm{tdil} (f_{n}^{-1}) \leq \frac{1}{1 - c_{n}}$; then there exists a subsequence $(n_{k})_{k \in \mathbb{N}}$ and a isometry $f$ such that $f_{n_{k}}$ converges to $f$ pointwise.       
\end{theo}  We recall that $x \prec y$ means that $x$ is in the causal past of $y$ and $x \ll y$ indicates that $y$ is in the chronological future of $x$.  Note also that we did not specify that $\ll$ is a partial order relation induced by a metric tensor $g$ since this would unnecessarily complicate the notation.  It should be clear from the context by which metric the particular partial order relation is defined.  We shall also not denote the distinction between $d_{g}$ and $d_{h}$.    \\*
\textsl{Proof}: \\*
Let $\mathcal{C}$ be a countable dense subset of $\mathcal{M}$.  By a diagonalization argument, one obtains a subsequence $\left\{ f_{n_{k}} \right\}$ such that $f_{n_{k}} ( p ) \stackrel{k \rightarrow \infty}{\rightarrow} f(p) \quad \forall p \in \mathcal{C}$.
Let $r$ be any interior point of $\mathcal{M}$ which is not in $\mathcal{C}$, we show now that the definition of $f$ can be extended to $r$ such that $\lim_{k \rightarrow \infty} f_{n_{k}} ( r ) = f(r)$.  Let $\mathcal{U}$ be a causally convex normal neighborhood of $r$ and choose a point $p_{1} \in \mathcal{C} \cap I^{-}(r) \cap \mathcal{U}$ close enough to $r$.  Let $\gamma$ be the unique timelike geodesic from $p_{1}$ through $r$ and define $\tilde{p}_{i} , \tilde{q}_{i} \in \gamma $ by $d(\tilde{p}_{i}, r) = \frac{d(p_{1},r)}{i} $ and $ d(r, \tilde{q}_{i}) = \frac{d(p_{1},r)}{i}$.  Hence $d( \tilde{p}_{i}, \tilde{p}_{i+j}) = \frac{j d(p_{1},r) }{i(i+j)} = d( \tilde{q}_{i+j}, \tilde{q}_{i})$ and $ d(\tilde{p}_{i},\tilde{q}_{i}) = \frac{ 2 d(p_{1},r)}{i} $.  Define now sequences $p_{i} , q_{i} \in \mathcal{C}$ such that $\tilde{p}_{i} \ll p_{i} \ll \tilde{p}_{i+1}$, $ \tilde{q}_{i+1} \ll q_{i} \ll \tilde{q}_{i}$ with the exception that $p_{1} = \tilde{p}_{1}$.  We shall now prove the following claims:
\begin{itemize}
\item  the sequence $( f(p_{i}))_{i \in \mathbb{N}_{0}}$ is contained in the compact set $A(f(\tilde{p}_{1}) , f(q_{1}))$\footnote{$A(p,q) = \{ r | p \leq r \leq q \}$.} and has exactly one accumulation point $f_{\uparrow} (r)$ which turns out to be a limit point. 
\item $f_{\uparrow} (r)$ is independent of the choice of $(p_{i})_{i \in \mathbb{N}}$
\end{itemize}
The first claim is an easy consequence of the observation that for all $i$ one has that :

\begin{eqnarray*}
d(f(\tilde{p}_{1}) , f(p_{i})) & = & \lim_{k \rightarrow \infty} d(f_{n_{k}} ( \tilde{p}_{1}) , f_{n_{k}} (p_{i})) \\
& = & d(\tilde{p}_{1} ,p_{i}), \\
\end{eqnarray*}
where we used the continuity of $d$ in the target space and the property of the convergence of the timelike dilatation and co-dilatation of the mappings $f_{n}$.  The above also proves that $ d(f(p_{i}) , f(p_{i+j})) = d(p_{i}, p_{i+j})$ and hence $ f(p_{i}) \ll f(p_{i+j})$.  This in turn implies that any accumulation point of the sequence $( f(p_{i}))_{i \in \mathbb{N}_{0}}$ must lie to the future of all $ f(p_{i})$, hence it is a limit point which must be unique.  \\*
The second claim follows from the observation that if $\hat{p}_{i}$ is another such sequence with corresponding $\hat{f}_{\uparrow} (r)$ then one has that $$p_{i} \ll \hat{p}_{i+1} \ll p_{i+2} \ll  \hat{p}_{i+3} \ll \ldots \ll r. $$  
Hence $$0 < d(f(\hat{p}_{i+1}), f_{\uparrow} (r) ) \leq d(f(p_{i}), f_{\uparrow} (r) ) - d(f(p_{i}),f(\hat{p}_{i+1})) $$ 
However, the first term on the rhs.\ converges to zero for $i \rightarrow \infty$ and the second term is estimated by $d(f(p_{i}),f(\hat{p}_{i+1})) \leq \frac{2 d(\tilde{p}_{1},r)}{i(i+2)}$.  Hence $\hat{f}_{\uparrow} (r) \in E^{-} ( f_{\uparrow} (r) )$.  The reverse is proven similary and this concludes the second claim. \\*
\\*  The same result is of course also true for $p$ replaced by $q$, and we denote the corresponding accumulation point by $f_{\downarrow} (r)$.   Note that $d(f(\tilde{p}_{1}),f_{\uparrow}(r)) = d(\tilde{p}_{1},r)$, $d(f_{\downarrow}(r), f(q_{1})) = d(r, q_{1})$ and $ d(f(\tilde{p}_{1}), f(q_{1})) = d(\tilde{p}_{1},q_{1})$, which all follow from continuity of $d$ and the present properties of $f$.   $f_{\uparrow}(r) = f_{\downarrow}(r)$ follows from the observation that changing $q_{1}$ by a point in the future of $q_{1}$, so that we can come arbitrary close to $\tilde{q}_{1}$, does not change the point $f_{\downarrow}(r)$.  For the same reasons as before, such a sequence of points $q_{1}$ will define a sequence $f(q_{1})$ which converges to a point, say, $f_{\uparrow}(\tilde{q}_{1})$ in the future of all points $f(q_{1})$.  Hence due to continuity we have that $d(f_{\downarrow}(r), f_{\uparrow}(\tilde{q}_{1})) = d(r, \tilde{q}_{1})$ and $d(f(\tilde{p}_{1}), f_{\uparrow}(\tilde{q}_{1})) = d(\tilde{p}_{1}, \tilde{q}_{1})$.  But this implies that $d(f_{\downarrow}(r),f_{\uparrow}(r)) = 0$ and more strongly $f_{\uparrow}(r) = f_{\downarrow}(r)$ otherwise by ``rounding off the edges'', we could find a timelike curve with length larger than $d(f(\tilde{p}_{1}), f_{\uparrow}(\tilde{q}_{1}))$, which is a contradiction.  It is now easy to see that $f_{n_{k}} (r) $ converges to $f(r)$, since for every $i$ we can find a $k_{0}$ such that for all $k \geq k_{0}$ one has that $f(p_{i}) \ll f_{n_{k}} ( p_{i+1} ) \ll f(r) \ll f_{n_{k}}(q_{i+1}) \ll f(q_{i}) $, which implies (because of the properties of $f_{n_{k}}$) that $f(p_{i}) \ll f_{n_{k}}(r) \ll f(q_{i}) $.  This concludes the proof when $r$ is an interior point, since the open Alexandrov sets $int(A(f(p_{i}),f(q_{i})))$ form a basis for the topology around $f(r)$.
The case when $r$ is a past boundary point is rather different, since then we cannot squeeze the point $r$ anymore in an Alexandrov set (the case of a future boundary point is identical).  Obviously, $f_{n_{k}}(r)$ belongs to the past boundary of $\mathcal{N}$.  Let $\gamma$ be the unique geodesic segment orthogonal to the past boundary in $r$, and choose the sequences $(\tilde{q}_{i})_{i \in \mathbb{N}}$ and $(q_{i})_{i \in \mathbb{N}}$ as before.  Then, we can find a subsequence $f_{n_{k_{l}}}$ such that $f_{n_{k_{l}}}(r) \stackrel{l \rightarrow \infty}{\rightarrow} f(r)$, where $f(r)$ belongs to the past boundary and $f_{n_{k_{l}}} (\gamma_{|\left[ r, \tilde{q}_{1}\right]}) \rightarrow f(\gamma_{|\left[ r, \tilde{q}_{1}\right]})$ in the $C^{0}$ topology of curves.  It is easy to see that $f(\gamma_{|\left[ r, \tilde{q}_{1}\right]})$ is the unique geodesic segment in $\mathcal{N}$ orthogonal to the past boundary in  $f(r)$.  But in this case, we have that $$f(q_{i}) \gg f_{n_{k}}(q_{i+1}) \gg f_{n_{k}}(r),$$ and since the $I^{-}(f(q_{i}))$ form a basis for the topology around $f(r)$, we have that $\lim_{k \rightarrow \infty} f_{n_{k}}(r) = f(r)$, which concludes the proof.  It is not difficult to see that $f$ is continuous by construction.  As a matter of fact, we should still prove that $f$ is onto.  Performing the same construction for $f_{n_{k}}^{-1}$ we find (by eventually taking a subsequence) a limit mapping $f^{-1}$.  We now    
show that $f^{-1} \circ f = id_{\mathcal{M}}$, $f \circ f^{-1} = id_{\mathcal{N}}$.  We shall prove the former, the proof of the latter is identical.  Suppose there exists an interior point $x$ such that $\lim_{k \rightarrow \infty} f_{n_{k}}^{-1} \circ f (x)  \neq x$, then there exist points $p_{1} , p_{2} , p_{3} , q_{1}, q_{2} , q_{3}$ such that 
$$ p_{1} \ll p_{2} \ll p_{3} \ll f^{-1} \circ f(x) \ll q_{3} \ll q_{2} \ll q_{1}$$ and $x \notin A(p_{1} , q_{1})$.  Then for $k$ big enough:
\begin{itemize}
\item $p_{3} \ll f_{n_{k}}^{-1} \circ f(x) \ll q_{3}$
\item $f_{n_{k}} ( p_{1} ) \ll f(p_{2}) \ll f_{n_{k}} ( p_{3})$
\item $f_{n_{k}}(q_{3}) \ll f(q_{2}) \ll f_{n_{k}} ( q_{1}), $
\end{itemize}    
hence 
$$ f_{n_{k}} (p_{1}) \ll f(p_{2}) \ll f(x) \ll f(q_{2}) \ll f_{n_{k}} ( q_{1}), $$ but $f_{n_{k}}(x) \notin A(f_{n_{k}} (p_{1}) , f_{n_{k}} (q_{1}))$, which implies that $f(x)$ cannot lie between $f(p_{2})$ and $f(q_{2})$, which is a contradiction.  Hence $f^{-1} \circ f$ equals the identity on the interior of $\mathcal{M}$, and therefore it equals the identity everywhere since it is continuous. The conclusion that $f$ is an isometry follows from the discussion in appendix A.  
 $\square$             
\\*
\\*
Remark first that in the proof of the theorem we needed the requirement that $\bigcup_{n} \left\{ f^{-1}_{n} (y) \right\}$ is precompact in $\mathcal{M}$ for all $y \in \mathcal{N}$ only to guarantee the surjectivity and hence the smoothness of $f$. Sensible questions are the following:
\begin{itemize}
\item Is $f$ not surjective a priori? If not give a counterexample.
\item Is the convergence uniform on compact sets with respect to some ``Riemannian'' metric $\tilde{d}$ on $\mathcal{N}$?  In either case, is the family of mappings $\left\{ f_{n} \right\}$ equicontinuous with respect to $\tilde{d}$? 
\item Give a counterexample to the conclusion of Ascoli-Arzela in case the spacetimes are not globally hyperbolic, but, say, causally continuous.  One might expect that such counterexample exists since we made use of \emph{all} properties of global hyperbolicity, that is the compactness of the Alexandrov sets to guarantee the convergence of the sequence $(f(p_{i}))_{i \in \mathbb{N}}$ and the continuity of $d$. 
\item Can one extend the theorem to the case where the timelike dilatation and co-dilatation of the mappings are only bounded and not necesserily convergent to $1$?
\end{itemize}  
I shall only examine the last question.  Remark first that the proof made crucial use of the fact that one has convergence to isometry.  The key argument was that the continuous timelike extension of $f$ maps $3$ points on a distance maximizing geodesic in $\mathcal{M}$ to $3$ points on a distance maximizing geodesic in $\mathcal{N}$.  This argument will clearly not be valid anymore when the limit mapping (if it exists) is not an isometry.  One could however invoke earlier the construction of $f^{-1}$ in the proof which is not a priori preferable considering the above questions.  However, such a strategy leads towards the following stronger result:
\begin{theo}
Let $\alpha < 1 < \beta , \quad f_{n} : \mathcal{M} \rightarrow \mathcal{N}$ be as in Theorem $2$ with the difference that $\textrm{tdil}(f_{n}) \leq \beta$ and $\textrm{tdil}(f_{n}^{-1}) \leq \frac{1}{\alpha}$.  Then there exists a subsequence $f_{n_{k}}$ and an $f$ such that $f_{n_{k}}$ converges pointwise to $f$.  Moreover one has that $\textrm{tdil}(f) \leq \beta$ and $\textrm{tdil}(f^{-1}) \leq \frac{1}{\alpha}$.   
\end{theo}
\textsl{Proof}: \\*
Let $\mathcal{C}$ and $\mathcal{D}$ be countable dense subsets in $\mathcal{M}$ and $\mathcal{N}$ respectively.  By a diagonalization argument we find a subsequence $f_{n_{k}}$ such that $f_{n_{k}} ( p )$ converges to $f(p)$ and $f_{n_{k}}^{-1}(q)$ converges to $f^{-1}(q)$ for all $p \in \mathcal{C}$ and $q \in \mathcal{D}$ respectively.
Suppose $r$ is an interior point and let $\gamma$, $(p_{i})_{i \in \mathbb{N}}$ and $(\tilde{p})_{i \in \mathbb{N}}$ be as before.  Take $q \in \mathcal{D}$ arbitrarily close in the chronological future of $f_{\uparrow}(r)$, we have then that     \begin{eqnarray*}
d(r,f^{-1}(q)) & = & \lim_{i \rightarrow \infty} d(p_{i}, f^{-1} (q) ) \\
& = & \lim_{i \rightarrow \infty} \lim_{k \rightarrow \infty} d(p_{i}, f_{n_{k}}^{-1} (q) ) \\
& \geq & \frac{1}{\beta} \lim_{i \rightarrow \infty} \lim_{k \rightarrow \infty} d(f_{n_{k}} (p_{i}), q) \\  
& \geq & \frac{d(f_{\uparrow}(r), q)}{\beta}. \\
\end{eqnarray*} 
Hence $r \ll f^{-1}(q)$.  Take now $q_{1}, q_{2} \in  \mathcal{D}$ such that $f_{\uparrow}(r) \ll q_{1} \ll q_{2}$ with $q_{2}$ arbitrarily close to $f_{\uparrow}(r)$.  Choose $i>0$, then for $k$ sufficiently large one has
$$ r \ll f_{n_{k}}^{-1}(q_{1}) \ll f^{-1} (q_{2}) $$
and
$$ f(p_{i}) \ll f_{n_{k}} ( p_{i+1} ) \ll f_{\uparrow}(r). $$
Hence 
$$ f(p_{i}) \ll f_{n_{k}}( p_{i+1}) \ll f_{n_{k}} ( r ) \ll q_{1}, $$
which proves $\lim_{k \rightarrow \infty} f_{n_{k}} (r) = f_{\uparrow}(r)$.  \\*
Let $r$ be a point of the ``past'' boundary (the future situation is dealt with identically).  Let $\gamma$ be a distance maximizing geodesic with past endpoint $r$ and let $(\tilde{q}_{i})_{i \in \mathbb{N}}, (q_{i})_{i \in \mathbb{N}}$ be sequences of points as before where now the ``futuremost'' point $\tilde{q}_{1}$ is sufficiently close to $r$ and $q_{1}$ can be chosen equal to $\tilde{q}_{1}$.  Without loss of generality, we can assume that $\tilde{q}_{1} \ll q \in \mathcal{D}$ such that $J^{-} (q)$ is compact.  For $k$ sufficiently large we have that
$$ f_{n_{k}} ( q ) \gg f(q_{1}) \gg f(q_{2}) \gg \ldots $$
Since $f_{n_{k}}$ is continuous $J^{-}(f_{n_{k}}(q)) = f_{n_{k}} ( J^{-} (
q))$ is compact, therefore the sequence $(f(p_{i}))_{i \in \mathbb{N}}$ has an accumulation point $f_{\downarrow}(r)$, which is as usual also a limit point.  Suppose $f_{\downarrow}(r)$ is not on the past boundary, then we can find a point $p \in \mathcal{D}$ such that $ p \ll f_{\downarrow} (r)$.  The calculation above shows that $f^{-1}(p) \ll r$, which is impossible.  Hence $f_{\downarrow}(r)$ belongs to the past boundary.  Since the past light cones $I^{-}(f(p_{i}))$ constitute a local basis for the topology around $f_{\downarrow}(r)$, the result follows.  
\\* 
The other conclusions of the theorem are obvious. $\square$ \\*
\\*
Having this theorem in the pocket, the theorem which guarantees convergence to isometry follows immediately.
\begin{theo}
Let $(\mathcal{M},g)$ and $(\mathcal{N},h)$ be compact globally hyperbolic spacetimes with boundary, then $d_{L} ((\mathcal{M},g),(\mathcal{N},h)) = 0$ iff $(\mathcal{M},g)$ and $(\mathcal{N},h)$ are isometric. 
\end{theo} 
The notion of Lipschitz distance however is too severe and does not give rise to a rich comparison theory since there is too much geometric control.  A result of Defrise-Carter ~\cite{Defrise} shows that every Lie algebra $\mathcal{L}$\footnote{The assumption in the paper of Defrise-Carter that the group needs to be finite dimensional, is not necessary.} of the group of \emph{local} conformal isometries of four dimensional Lorentz manifolds are all, with two exceptions, essentially isometries.  By this, I mean that for every spacetime $(\mathcal{M},g)$ not conformally equivalent to Minkowski or a plane-wave spacetime with parallel rays, there exists a conformal factor $\Omega$ such that $\mathcal{L}$ constitutes a $r$ dimensional Lie algebra of isometries for the spacetime $(\mathcal{M}, \Omega g)$, with $r \leq  10$.  In Minkowski spacetime,  there is a $15$-dimensional group of proper conformal transformations\footnote{Generators consist of the $10$ Poincare transformations, $1$ dilatation and $4$ accelerations.}  and in the latter only a $6$ or $7$ dimensional group of homotheties\footnote{$5$ respectively $6$ generators form an isometry group, and $1$ generator forms a dilatation.}.  Hence, there are ``not many'' infinitesimal conformal isometries, and there are even fewer which can be integrated.  Note that the result of Defrise-Carter does not mention anything about \emph{discrete} conformal isometries.  However, the results of this section are still very important, since:    
\begin{itemize}
\item we shall be forced to generalize this Lipschitz theory to \emph{abstract} globally hyperbolic Lorentz spaces, which will be done in the next paper.    
\item the proofs give a hunch how to prove convergence to isometry in case the family of mappings gets enlarged, such as will happen in the next section.
\end{itemize}
In a complete Riemannian manifold which is not locally flat, Kobayashi and Numizu have proven that there are no homotheties which are not isometries.  As stated before, this is not true in the Lorentzian case as the next plane wave spacetime shows \cite{Beem}.  \\*
\\*
\textbf{Example}
Consider $\mathbb{R}^{3}$ with the metric $ds^{2} = e^{xz}dxdy + dz^{2}$.  $(\mathbb{R}^{3}, ds^{2})$ is not flat and the mappings $\phi_{t}(x,y,z) = (e^{t}x,e^{-3t}y,e^{-t}z)$ are proper homothecies with factor $e^{-2t}$.  \\*
\\*
Moreover, we will see that for every compact globally hyperbolic spacetime $(\mathcal{M},g)$ there exists a ``Riemannian'' metric $D_{\mathcal{M}}$ such that all $d_{g}$ isometries are $D_{\mathcal{M}}$ isometries.  Suppose $(\mathcal{M},g)$ is not a compact piece cut out of Minkowski or a plane-wave spacetime (with parallel rays), then there exists a ``Riemannian'' metric $\tilde{D}_{\mathcal{M}}$ such that ``most'' (apart from eventual discrete conformal isometries) $g$-conformal isometries are $\tilde{D}_{\mathcal{M}}$ isometries\footnote{We know there exists a global conformal factor $\Omega$ such that essentially all $g$ conformal isometries are $\Omega g$ isometries, hence the claim follows.}.      
\section{A Gromov-Hausdorff distance} 
As in the previous section, we recall the notion of Gromov-Hausdorff distance in the ``Riemannian'' case.  For this purpose define the Hausdorff distance $d_{H}$ between subsets $U,V$ of a metric space $(X,d_{X})$ as 
$$ d_{H} (U,V) = \inf \{\epsilon | U \subset B(V, \epsilon), V \subset B(U, \epsilon) \}$$
where $B(U, \epsilon) = \{ x \in X | \exists a \in U : d_{X}(x,a) < \epsilon \}$. 
Gromov had around $1980$ the following idea \cite{Gromov} :  consider two compact metric spaces $(X,d_{X})$ and
$ ( Y, d_{Y})$, define a metric $d$ on the disjoint union $X \cup Y$ to be \emph{admissible} iff the restrictions of $d$ to $X$ and $Y$ equal $d_{X}$ and $d_{Y}$ respectively.  Then $$d_{GH} ( (X,d_{X}),(Y,d_{Y})) = \inf \{ d_{H} (X,Y) | \textrm{all admissible metrics on} X \cup Y \}.$$
In other words the Gromov-Hausdorff distance between two metric spaces is the infimum over all Hausdorff distances in 
$X \cup Y$ with respect to metrics which extend the given metrics on $X$ and $Y$.  Suppose $d$ is an admissible metric on $X \cup Y$; then there exist mappings $f : X \rightarrow Y$, $ g :Y  \rightarrow X$ such that $d (x,f(x)) \leq d_{H}(X,Y)$ and $d(y,g(y)) \leq d_{H}(X,Y)$ for all $x \in X$, $y \in Y$ respectively. 
The triangle inequality and the properties of $d$ imply that :
\begin{eqnarray}
\left| d_{Y} ( f(x_{1}) , f(x_{2})) - d_{X} ( x_{1} , x_{2} ) \right| & \leq & 2 d_{H}(X,Y)  \\
\left| d_{X} ( g(y_{1}) , g(y_{2})) - d_{Y} ( y_{1} , y_{2} ) \right| & \leq & 2 d_{H}(X,Y)  \\ 
d_{X} ( x , g \circ f (x) ) & \leq & 2 d_{H}(X,Y) \\
d_{Y} (y , f \circ g (y )) & \leq & 2 d_{H}(X,Y) 
\end{eqnarray} 
Observe that the last two inequalities imply that in the limit for $d_{H} (X , Y)$ to zero, $f$ becomes invertible.  But for compact metric spaces, invertibility also follows from the observation that in the limit for $d_{H} (X , Y)$ to zero, $f$ and $g$ become distance-preserving maps.  Hence $g \circ f$ and $f \circ g$ are distance-preserving maps on $X$ and $Y$ respectively.  The compactness assumption then implies that they are both bijections and, as a consequence, so are $f$ and $g$.  We shall first prove a similar result in the Lorentzian case.
\begin{theo}
Let $f : \mathcal{M} \rightarrow \mathcal{M}$ be continuous and Lorentzian distance preserving on the interior of $\mathcal{M}$; then $f$ maps the interior onto itself.
\end{theo}  
\textsl{Proof}:
Remark that an interior point is mapped by a distance-preserving map to an interior point.  Suppose $p$ is an interior point not in $f(\mathcal{M})$, then there exists a neighborhood $\mathcal{U}$ of $p$ for which $f(\mathcal{M}) \cap \mathcal{U} = \emptyset$.  For suppose not, then we can find a sequence $r_{n} \stackrel{n \rightarrow \infty}{\rightarrow} r$ such that $f(r_{n}) \stackrel{n \rightarrow \infty}{\rightarrow} p$.  Hence $r$ is not an interior point and without loss of generality we can assume it belongs to the future boundary.  But then $f(\mathcal{M}) \cap I^{+} ( p) = \emptyset$, otherwise there would exist an interior point to the future of all $r_{n}$, which is impossible.\\*
Hence, we may assume that there exist points $r \ll p \ll s$ such that $f(\mathcal{M}) \cap I^{+}(r) \cap I^{-}(s) = \emptyset$ and $d_{g}(r,p) = d_{g} ( p,s) > 0$.   Since $f^{k} (p) \notin I^{+}(r) \cap I^{-}(s)$ for all $k$, we get that $f^{k}(p) \notin  I^{+}(f^{l}(r)) \cap I^{-}(f^{l}(s))$ for all $k \geq l$.  By taking a subsequence if necessary, we can assume that $f^{n} ( p) \stackrel{n \rightarrow \infty}{\rightarrow} \tilde{p}$, $f^{n} ( r) \stackrel{n \rightarrow \infty}{\rightarrow} \tilde{r}$, $f^{n} ( s) \stackrel{n \rightarrow \infty}{\rightarrow} \tilde{s}$.  Hence $\tilde{r} \ll \tilde{p} \ll \tilde{s}$, but this is impossible since this implies that for $n$ big enough $\tilde{p} \in I^{+}(f^{n}(r)) \cap I^{-}(f^{n}(s))$.  $\square$ \\*
\\*     
Let us now make the following definition.   
\begin{deffie}
(\textbf{Lorentzian Gromov-Hausdorff} ) We call $(\mathcal{M},g)$ and $(\mathcal{N},h)$ $\epsilon$-close iff there exist mappings $\psi : \mathcal{M} \rightarrow \mathcal{N}$, $\zeta : \mathcal{N} \rightarrow \mathcal{M}$ such that 
\begin{eqnarray}
\left| d_{h} ( \psi ( p_{1} ), \psi (p_{2})) - d_{g} (p_{1} , p_{2} ) \right| & \leq & \epsilon \quad \forall p_{1}, p_{2} \in \mathcal{M} \\
\left| d_{g} ( \zeta ( q_{1} ) ,\zeta (q_{2})) - d_{h} ( q_{1} ,q_{2} )\right| & \leq & \epsilon \quad \forall q_{1}, q_{2} \in \mathcal{N}. 
\end{eqnarray}
The Gromov-Hausdorff distance $d_{GH}((\mathcal{M},g),(\mathcal{N},h))$ is defined as the infimum over all $\epsilon$ such that $(\mathcal{M},g)$ and $(\mathcal{N},h)$ are $\epsilon$-close.
 
\end{deffie}
Suppose we are given sequences $(\psi_{n})_{n \in \mathbb{N}}, (\zeta_{n})_{n \in \mathbb{N}}$ of -possibly discontinuous - maps which make $(\mathcal{M},g)$ and $(\mathcal{N},h)$ $\frac{1}{n}$ close.  Then, because of the previous theorem, any limit mapping is necessarily an isometry.
\begin{theo}
$d_{GH}((\mathcal{M},g),(\mathcal{N},h)) = 0$ iff $(\mathcal
{M},g)$ and $(\mathcal{N},h)$ are isometric.
\end{theo}     
\textsl{Proof}: \\*
Let $\mathcal{C}$ and $\mathcal{D}$ be countable dense subsets of $\mathcal{M}$ respectively $\mathcal{N}$, and take subsequences $(\psi_{n_{k}})_{k \in \mathbb{N}}$ and $(\zeta_{n_{k}})_{k \in \mathbb{N}}$ such that 
\begin{itemize}
\item $\psi_{n_{k}} ( p ) \stackrel{k \rightarrow \infty}{\rightarrow} \psi(p)$ for all $p \in \mathcal{C}$ 
\item $\zeta_{n_{k}} ( q ) \stackrel{k \rightarrow \infty}{\rightarrow} \zeta(q)$ for all $q \in \mathcal{D}$ 

\end{itemize}
Obviously $d_{h} ( \psi ( p ) , \psi ( \tilde{p} )) = d_{g} ( p , \tilde{p})$ for all $p, \tilde{p} \in \mathcal{C}$ and $ d_{g} ( \zeta ( q ) , \zeta ( \tilde{q} )) = d_{h} ( q , \tilde{q})$ for all $q, \tilde{q} \in \mathcal{D}$, which is an easy consequence of the global hyperbolicity and the limiting properties of the sequences $(\psi_{n_{k}})_{k \in \mathbb{N}}$ and $(\zeta_{n_{k}})_{k \in \mathbb{N}}$. \\*
We shall now prove that the limit map $\psi$ exists and is distance-preserving.
Let $r$ be an interior point of $\mathcal{M}$ and take sequences $(\tilde{p}_{i})_{i \in \mathbb{N}}$ , $(\tilde{q}_{i})_{i \in \mathbb{N}}$, $(p_{i})_{i \in \mathbb{N}}$ and $(q_{i})_{i \in \mathbb{N}}$ in $\mathcal{M}$ as before.  
In exactly the same way as in the proof of theorem $2$, we obtain that $\psi_{ \uparrow } (r ) = \psi_{ \downarrow } (r)$.  Also $\psi(r) = \lim_{k \rightarrow \infty} \psi_{n_{k}} (r ) $ since for arbitrary $i$ we can find a $k_{0}$ such that $\forall k \geq k_{0}$ 
\begin{itemize}
\item $\frac{1}{k} < \min \{ d(p_{i + 1},r) , d(r , q_{i+1}) \}$
\item $\psi(p_{i}) \ll \psi_{n_{k}} ( p _{i+1} ) \ll \psi_{n_{k}} ( q_{i+1} ) \ll \psi (q_{i} )$ 
\end{itemize}  
hence 
$$ \psi (p_{i}) \ll \psi_{n_{k}} ( p_{i+1} ) \ll \psi_{n_{k}} (r) \ll \psi_{n_{k}} ( q_{i+1} ) \ll \psi ( q_{i} ) $$
which proves the case.  From this it is easy to prove that $\psi$ is continuous on the interior points. \\*
In exactly the same way one constructs a continuous limit map $\zeta$ on the interior of $\mathcal{N}$.  \\*
The previous theorem now shows that $\psi$ and $\zeta$ are distance preserving homeomorphisms from the interior of $\mathcal{M}$ to $\mathcal{N}$ and from the interior of $\mathcal{N}$ to $\mathcal{M}$ respectively.  Using this, it is not difficult to show that one can continuously extend $\psi$ to the boundary so that $\lim_{k \rightarrow \infty} \psi_{n_{k}} ( r ) = \psi(r)$ for every boundary point $r$.  Hence the result follows.$\square$   \\*
\\*
Furthermore, it is obvious that $d_{GH}$ is symmetric and satisfies the triangle inequality.  We will now discuss some properties of $d_{GH}$.  Let us start with an obvious one which is similar to the ``Riemannian'' case.  
\begin{theo} \,
$d_{GH} ((\mathcal{M},g),(\mathcal{N},h)) \leq \max \{\textrm{tdiam}(\mathcal{M}), \textrm{tdiam} (\mathcal{N}) \}$ where \\* $\textrm{tdiam}( \mathcal{M} )$ denotes the timelike diameter, i.e.,
$$ \textrm{tdiam}( \mathcal{M} ) = \max_{p,\tilde{p} \in \mathcal{M}} d_{g} ( p ,\tilde{p} ).$$
\end{theo}  
We shall now give an example that might feel strange in the beginning for people used to Riemannian geometry, although the result itself is what one should expect from Lorentzian geometry.  \\* 
\\*
\textbf{Example}
\\*
As mentioned before, a need will present itself for abstraction of the concept of Lorentzian manifold.  Therefore, it is not hard to imagine that a Riemannian manifold is a Lorentz space where every point is null-connected with itself and not causally related to any other point (imagine that the Riemannian manifold serves as a spacelike Cauchy surface in a globally hyperbolic spacetime).  The previous theorem shows then that \emph{any} two Riemannian manifolds are a distance zero apart since their timelike diameters are zero.  This is very much different from the usual ``Riemannian'' theory \emph{but} in a purely Lorentzian theory this result is obvious from the fact that the causal distance does not provide us with any information whatsoever.  Hence the manifold would be unobservable, so how could one compare two things which cannot be observed?  This result shows that, if one wants the moduli space to be a complete metric space, the timelike diameter needs to be controlled, i.e., bounded away from zero as the next example shows.  \\* Consider cylinders $C_{T} = S^{1} \times \left[ 0 , T \right]$ with Lorentz metric $ds^{2} = - dt^{2} + d \theta^{2}$.  A Gromov-Hausdorff limit for $T \rightarrow 0$ is $S^{1}$, but it could equally well be any other Riemannian manifold. 
$\square$    
\\*
\\*
Now, we shall show that we can construct a metric $D_{\mathcal{M}}$ such that every $d_{g}$ isometry is a $D_{\mathcal{M}}$ isometry.  This metric shall be constructed from the Lorentzian distance $d_{g}$ alone, which is in contrast to the usual extra assumption of a preferred class of observers in the major part of the literature.  Such a preferred class of observers is for example given if the energy momentum tensor satisfies the type I weak energy condition ~\cite{Hawking1}, i.e., determines a preferred timelike eigenvectorfield.  However, our approach is purely geometrical and matter is not assumed to determine geometry through the Einstein equations.  This is moreover the only sensible strategy if 
\begin{itemize}
\item one wants to construct a theory of \emph{vacuum} quantum gravity
\item one considers spacetime \emph{not} to be a manifold.  What would the analogue be of the Einstein-Hilbert action on something like a causal set ~\cite{Sorkin} or a spin network ~\cite{Thiemann}, anyway?
\end{itemize}         
$D_{\mathcal{M}}$ will also play a crucial part in the construction of the limit space of a Cauchy sequence of compact interpolating spacetimes \cite{Noldus1}.  For reasons which will become clear in \cite{Noldus1}, $D_{\mathcal{M}}$ will be refferred to as the strong metric.  
\begin{deffie} 
Let $(\mathcal{M},g)$ be a compact interpolating spacetime, the strong  metric $D_{\mathcal{M}}$ is defined as
$$ D_{\mathcal{M}} ( p , q ) = \max_{r \in \mathcal{M}} \left| d(p,r) + d(r,p) - d(q,r) - d(r,q) \right| $$
$\square$
\end{deffie}
\textbf{Note:} The reader should note that the strong metric could be defined on any set (with max replaced by sup) equipped with a Lorentz distance.  This remark will lead to the notion of Lorentz space \cite{Noldus2}.  $\square$
\\*
\\*
We end this section with a theorem which is an amalgamation of elementary properties of the strong metric.
\begin{theo}  This theorem is an amalgamation of results concerning the strong metric.
\begin{itemize}
\item{a)} $d_{g}$ is continuous in the strong topology. 
\item{b)} The Alexandrov topology is weaker than the strong topology.   
\item{c)} On a compact globally hyperbolic spacetime $\mathcal{M}$, the manifold, strong, and Alexandrov topology coincide.   
\item{d)} The $\epsilon$-balls of the metric $D_{\mathcal{M}}$ are causally convex, i.e., if $p \ll q$ and $p,q \in B(r, \epsilon)$ for some $r \in \mathcal{M}$, then $I^{+}(p ) \cap I^{-}(q) \subset B(r , \epsilon)$.  
\end{itemize}
$\square$
\end{theo}      
\textsl{Proof}:  \\*
\begin{itemize}
\item{a)} Choose $p,q \in \mathcal{M}$, $\epsilon > 0$, $r \in B( p , \frac{\epsilon}{2})$, $s \in B( q , \frac{ \epsilon}{2} )$; then
$$ \left| d(p,q) - d(r,s) \right| \leq \left| d(p,q) - d(p,s) \right| + \left| d(p,s) - d(r,s) \right| < \epsilon $$ 
\item{b)} Since $d$ is continuous in the strong topology, $d(r,\cdot ),d( \cdot ,r): \mathcal{M} \rightarrow \mathbb{R}^{+}$, are also continuous in the strong topology for all $r \in \mathcal{M}$.  Hence $d(r, \cdot )^{-1} ( \left( 0 , +\infty \right) )$ and $ d( \cdot ,r)^{-1} (\left( 0 , +\infty \right) )$ are open in the strong topology, which implies that the Alexandrov topology is weaker than the strong one.
\item{c)} Since $D_{\mathcal{M}}$ is continuous in the manifold topology, the strong topology is weaker than the manifold topology.  But the Alexandrov topology is weaker than the strong topology and coincides with the manifold topology, hence all topologies coincide.
\item{d)} Follows from the definition.
\end{itemize}
$\square$  
\section{Acknowledgements}
I want to thank my promotor Norbert Van den Bergh and co-promotor Frans Cantrijn for their careful reading of the manuscript and the interest shown in my research.  Also I express my gratitude to Luca Bombelli and Denis Constales for proofreading this paper and providing me with useful comments.        
\section{Appendix A}
The next theorem is a \emph{slight} generalization of the result in \cite{Beem}.
\begin{theo}
If $(\mathcal{M},g)$ is strongly causal, then every onto map $f: \mathcal{M} \rightarrow \mathcal{N}$ with finite, strictly positive timelike dilatation and co-dilatation is a homeomorphism     
\end{theo}
\textsl{Proof} \\*

Observe first that for all $p,q \in \mathcal{M}$, one has that $d(f(p),f(q)) > 0$ iff $d(p,q)>0$.  Hence, $f(I^{ \pm}(x)) = I^{ \pm}(f(x))$ (since $f$ is onto) and $f(I^{+}(p) \cap I^{-}(q)) = I^{+}(f(p)) \cap I^{-} (f(q))$.  Since $(\mathcal{M},g)$ is strongly causal, the Alexandrov topology coincides with the manifold topology.  Hence, $f$ is an open mapping.  $f$ is also injective, since if $p \neq q$ and $f(p) = f(q)$, we arrive to the following contradiction.  Let $\mathcal{U}$ be a locally convex neighborhood of $p$ which does not contain $q$ and satisfies the condition that every causal curve intersects $\mathcal{U}$ exactly once.  Take then $r \ll p \ll s$ with $r,s \in \mathcal{U}$ then $I^{+} (r) \cap I^{-} (s) \subset \mathcal{U}$.  A fortiori $f(r) \ll f(p) = f(q) \ll f(s)$ which is a contradiction since $q \notin I^{+} (r) \cap I^{-} (s)$.  We are done if we prove that $f^{-1}$ is open. For this it is sufficient to prove that $(\mathcal{N},h)$ is strongly causal.  Suppose that strong causality is not satisfied at $f(p)$.  First, choose a locally convex neighborhood $\mathcal{U}$ of $f(p)$ such that $(\mathcal{U},h_{ | \mathcal{U} })$ is globally hyperbolic. Let $\mathcal{W}$ be a neighborhood of $f(p)$ of compact closure in $\mathcal{U}$.  If strong causality is not satisfied at $f(p)$ then there exist points $q_{n} \ll f(p) \ll  r_{n}$ in $\mathcal{W}$ such that $q_{n},r_{n} \stackrel{n \rightarrow \infty}{\rightarrow} f(p)$ and causal curves $\lambda_{n}$ from $p_{n}$ to $q_{n}$ which leave $\mathcal{U}$.  Denote by $z_{n}$ the first intersection with $\partial \mathcal{W}$ of $\lambda_{n}$.  Then there exists a subsequence $z_{n_{k}}$ such that $z_{n_{k}} \stackrel{k \rightarrow \infty}{\rightarrow} z$.  Obviously, $p = f^{-1} (z)$ otherwise the continuity of $f^{-1}$ would contradict the strong causality of $(\mathcal{M},g)$.  But on the other hand $p = f^{-1} (z)$ contradicts the injectivity of $f$.  $\square$   \\*
We show now that $f$ takes null geodesics to null geodesics.  Take a small enough convex, normal neighborhood $\mathcal{U}$ of $p$ which no causal curve intersects more than once and such that $(\mathcal{U} , g_{| \mathcal{U} } )$ is globally hyperbolic.  Moreover, we assume that the closure of $f(\mathcal{U})$ belongs to a convex, normal neighborhood $\mathcal{V}$ of $f(p)$ which no causal curve intersects more than once, with $(\mathcal{V}, h_{| \mathcal{V}})$ globally hyperbolic.  Let $\alpha(q,r)$ be a null geodesic in $\mathcal{U}$ and take sequences $q_{n} \rightarrow q$, $r_{n} \rightarrow r$ with $q_{n} \ll r_{n}$ for all $n$.  $f$ takes timelike geodesics $\alpha(q_{n},r_{n})$ with length $d(q_{n} , r_{n} )$ to timelike curves $\gamma(f(q_{n}),f(r_{n}) )$  with length at most $\beta d(q_{n} ,r_{n})$.  Moreover, $f(q_{n}) \rightarrow f(q)$ and $f(r_{n}) \rightarrow f(r)$.  The geodesics $\alpha(q_{n} ,r_{n} )$ converge to the null geodesic $\alpha(q,r)$.  Because of the global hyperbolicity of $(\mathcal{V}, h_{| \mathcal{V}})$ a subsequence of the timelike curves $\gamma(f(q_{n}),f(r_{n}))$ converges to a causal curve from $q$ to $r$.  This causal curve need to be an unbroken null geodesic $\alpha(f(q),f(r))$ since $d(f(q),f(r)) = 0$.  In fact, it is easy to see that the whole sequence $\gamma(f(q_{n}) ,f( r_{n}))$ converges in the $C^{0}$ topology of curves to $\alpha(f(q),f(r))$, which concludes the proof.  \\*  It is easy to check that if $(\mathcal{M},g)$ is a strongly causal spacetime with spacelike boundary, then the above results are still valid, i.e., the homeomorphism extends to the boundary. 
A well known result of Hawking, King and McCarthy \cite{Hawking2}, which is the Lorentzian equivalent of an earlier theorem by Palais, states that every homeomorphism which maps null geodesics to null geodesics must be a conformal isometry.          \section{Appendix B}
In this appendix we sketch how to locally extend a conformal diffeomorphism across the boundary.  Let $r$ be a boundary point of $\mathcal{M}$ and choose a neighborhood $\mathcal{U}$ of $r$ diffeomorphic (under $\psi$) to the open hypercube union one side hyperplane $H$.  Let $(\mathcal{V}, g_{\mathcal{V}})$ be a local extension in $\mathbb{R}^{n}$ of $(H, \psi_{*}g)$.  Choose $(W, \phi_{*} h)$ to be a similar hypercube neighborhood of $f(r)$ and interpret $f$ as a conformal mapping of $(H,\psi_{*}g)$ to $(W, \phi_{*}h)$ with conformal factor $\Omega$.  Obviously $f$ can be $C^{\infty}$ locally extended over the boundary around $\psi (r) $, as can $\Omega$ such that $\Omega$ has almost vanishing normal first derivative \footnote{remember that we have chosen our coordinates in such a fashion that the boundary corresponds to a part of the hypersurface $t=0$.}.  Then we can construct an extension of $(W, \phi_{*}h)$ by defining the extension $\widetilde{\phi_{*}h}$ of $ \phi_{*}h$ around $\phi(f(p))$ as $\widetilde{\phi_{*}h} = \frac{(f \circ \psi)_{*}g}{\Omega}$.  Now the real question is the following: let $z$ be a conformal $C^{\infty}$ diffeomorphism from $(H, \psi_{*}g)$ to $(W, \phi_{*} h)$; does there exist an extension $\tilde{z}$ from $(\mathcal{V}, g_{\mathcal{V}})$ to $(f(\mathcal{V}), \widetilde{\phi_{*}h})$ ?  Clearly, if such a \emph{local} extension extension exists, it must be unique.  This can be seen as follows.  Take $x \in \mathcal{V} \cap H^{c}$ close enough to the boundary $t=0 \leftrightarrow \Sigma $ such that $E^{+}(x) \cap \Sigma$ is diffeomorphic to the $2$ - sphere and such that the null geodesics can be pushed over the boundary a bit (ie. there are no cut points in a neighborhood of $\Sigma$ wrt. $\psi_{*}g$ for $x$ sufficiently close to it).  Denote by $T_{\textrm{null}} \Sigma $ the bundle of null vectors over $\Sigma$, hence $x$ determines a unique (discontinuous) section $\rho_{x}$ with support in $E^{+}(x) \cap \Sigma$ such that
$$ \exp( - \rho_{x} (y) ) = x \quad \forall y \in E^{+}(x) \cap \Sigma .$$
The push forward under $z$ of the section $\rho_{x}$ determines uniquely the point $z(x)$ as the first past intersection point of the null rays defined by $z_{*} \rho_{x}$ - if it exists.  Now, since $f^{-1} \circ z$ is a conformal diffeomorphism if only if $z$ is, it is sufficient to prove the existence of the unique conformal extension of the former.  Clearly, in two dimensions, such intersection point exists and therefore also the conformal extension.  However, I have no argument for now which proves the result in dimension greater than $2$.

\end{document}